\documentclass[twocolumn,prb,amsmath,amssymb,floatfix,superscriptaddress]{revtex4-2}   
\usepackage{graphicx}
\usepackage{dcolumn}
\usepackage{bm}
\usepackage{xspace}
\usepackage{hyperref}
\usepackage{mathtools}
\usepackage{color}
\usepackage[flushleft]{threeparttable}

\def\nio{Na$_2$IrO$_3$\xspace}

\begin{document}
\title{Spectrum of the hole excitation in spin-orbit Mott insulator \nio}
\author{Wei~Wang}
\affiliation{School of Science, Nanjing University of Posts and Telecommunications, Nanjing 210023, China}
\author{Zhao-Yang~Dong}
\affiliation{Department of Applied Physics, Nanjing University of Science and Technology, Nanjing 210094, China}
\author{Shun-Li~Yu}
\email{slyu@nju.edu.cn}
\affiliation{National Laboratory of Solid State Microstructures and Department of Physics, Nanjing University, Nanjing 210093, China}
\affiliation{Collaborative Innovation Center of Advanced Microstructures, Nanjing University, Nanjing 210093, China}
\author{Jian-Xin Li}
\email{jxli@nju.edu.cn}
\affiliation{National Laboratory of Solid State Microstructures and Department of Physics, Nanjing University, Nanjing 210093, China}
\affiliation{Collaborative Innovation Center of Advanced Microstructures, Nanjing University, Nanjing 210093, China}


\begin{abstract}
We study the motion of a hole with internal degrees of freedom, introduced to the zigzag magnetic ground state of \nio, by using the self-consistent Born approximation. We find that the low, intermediate, and high-energy spectra are primarily attributed to the singlet, triplet, and quintet hole contributions, respectively. The spectral functions exhibit distinct features such as the electron-like dispersion of low-energy states near the $\Gamma$ point, the maximum M-point intensity of mid-energy states, and the hole-like dispersion of high-energy states. These features are robust and almost insensitive to the exchange model and Hund's coupling, and are in qualitative agreement with the angular-resolved photoemission spectra observed in \nio. Our results reveal that the interference between internal degrees of freedom in different sublattices plays an important role in inducing the complex dispersions.
\end{abstract}

\maketitle

The charge dynamics of doped Mott insulators have attracted significant attention due to their crucial role in understanding the high-temperature superconducting mechanism in high-$T_c$ cuprates \cite{RevModPhys.78.17,PhysRevLett.61.365}. In this regard, the $t$-$J$ model, featuring antiferromagnetic Heisenberg interactions, is widely considered as an appropriate model for describing the low-energy behavior of doped antiferromagnetic Mott insulator\cite{PhysRevB.37.3759}. In such an antiferromagnetically ordered background, the coherent propagation of a single hole at low energies has been theoretically established on the square lattice \cite{RevModPhys.78.17,PhysRevB.39.6880,PhysRevB.44.317,PhysRevB.44.2414,RevModPhys.66.763,PhysRevB.52.R15711} and verified by the angle-resolved photoemission spectroscopy (ARPES) experiments\cite{RevModPhys.75.473}.

Recently, much attention has been focused on the spin-orbit assisted Mott insulators, where the strong spin-orbit coupling gives rise to a low-energy $J=1/2$ (multi-electron state) Kramers doublet and a high-energy $J=3/2$ quadruplet from $t_{2g}^5$ manifold\cite{PhysRevLett.101.076402,science.1167106}, such as the honeycomb-lattice Na$_2$IrO$_3$ \cite{PhysRevB.82.064412,PhysRevLett.108.127203,Rau2016,
PhysRevLett.102.017205,PhysRevLett.117.187201,PhysRevLett.117.187201} and $\alpha$-RuCl$_3$\cite{PhysRevB.90.041112,Banerjee2016,Winter2017,Ran2017}. The effective $J=1/2$ spin-orbit entangled states on the honeycomb lattice\cite{PhysRevLett.102.017205} could induce bond-dependent Kitaev interactions, which can generate a spin-liquid ground state\cite{Kitaev2006}. However, zigzag magnetic ordering\cite{PhysRevLett.108.127203,PhysRevB.92.235119,Banerjee2016,PhysRevB.93.134423} has been experimentally observed in these materials at low temperatures, implying the existence of other exchange terms that stabilize the magnetic order. To describe the low-energy magnetic physics, several extended models have been proposed, ranging from nearest-neighbor Heisenberg terms\cite{PhysRevLett.102.017205,PhysRevLett.105.027204,PhysRevLett.110.097204} and off-diagonal $\Gamma$ interaction\cite{PhysRevLett.112.077204,PhysRevLett.113.107201,PhysRevB.93.214431,PhysRevB.93.155143,Wang2017} to longer-ranged Heisenberg\cite{PhysRevB.84.180407,PhysRevB.93.214431} and Kitaev\cite{Rousochatzakis2015} interactions. Based on the extended exchange models with hopping terms, which are called the one-band $t$-$J$-like model, theoretical studies\cite{Trousselet2013,Trousselet2014,Wang_2018} have shown that a single hole moves incoherently at high energies and hidden quasiparticles arise at low energies in the zigzag magnetic phase. In these works, the hole has only one degree of freedom. However, in real materials, a hole doped in Ir$^{4+}$ of \nio, i.e. removing an electron from $t^5_{2g}$ configuration, can be doped not only into $j=1/2$ (single-electron state) states but also into $j=3/2$ states. The multiple habitability of a hole makes the hole propagation more complicated than that in previous works\cite{Trousselet2013,Trousselet2014,Wang_2018}. Therefore,  the dynamics of a single hole doped into the $t^5_{2g}$ states in the zigzag phase on the honeycomb lattice deserves further studies.

Experimently, ARPES studies\cite{PhysRevLett.109.266406,Alidoust2016,Moreschini2017,PhysRevB.101.235415} have found various features in the low-energy electronic structure in \nio.
Comin \emph{et al.}\cite{PhysRevLett.109.266406} detected four flat bands and a clear insulating gap of $\sim340$ meV, which is explained by the interference between the two sublattices of the honeycomb lattice\cite{Trousselet2013}. Alidoust \emph{et al.}\cite{Alidoust2016} and Rodriguez \emph{et al.}\cite{PhysRevB.101.235415} reported the observation of metallic states at the center of the BZ. Moreschini \emph{et al.}\cite{Moreschini2017} found that the dispersion of the lowest-energy state near the $\Gamma$ point is electronlike and does not cross the Fermi level, and that there exists flat bands at $\sim0.7$ eV and dispersive holelike states at the higher-energy region. The intensity of the flat bands focuses on the M point, and that of the holelike states near $\sim1.2$ eV  is centered on the K point. The charge dynamics in \nio revealed by these experiments are much more complex than the previous theoretical results based on the one-band $t$-$J$-like models\cite{Trousselet2013,Trousselet2014,Wang_2018}.
Therefore, the understanding of the spectral characteristics unveiled by ARPESs in the spin-orbit assisted Mott insulator \nio remains an open question.

\begin{figure}
  \centering
  \includegraphics[width=0.45\textwidth, trim={0.4cm 0.0cm 0.5cm 0.1cm}, clip]{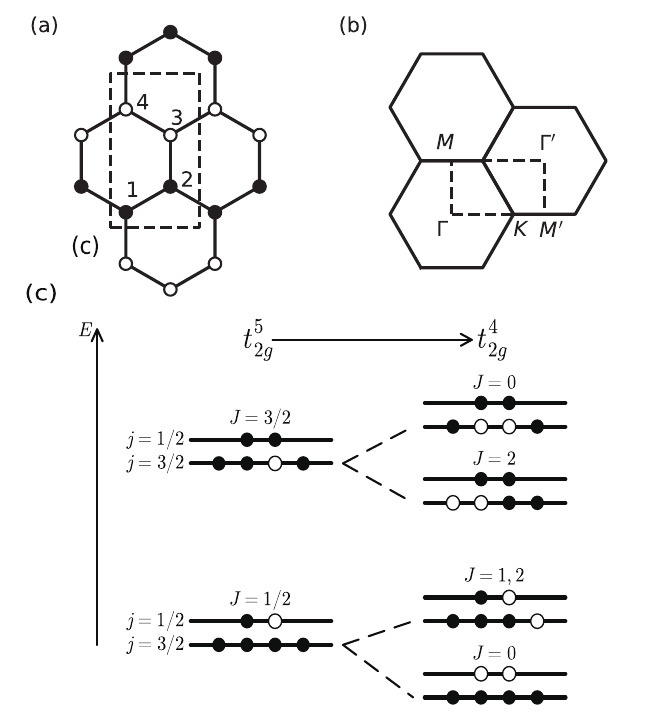}
  \caption{\label{fig1}(a) Magnetic structure of the zigzag ordering,  where the filled and open sites denote pseudospin up and down, respectively. The rectangle is magnetic unit cell. (b) Structure of the reciprocal space. $M$, $\Gamma$, $K$, $M^\prime$, and $\Gamma^\prime$ are the high-symmetry points. (c) States of $t_{2g}^5$ and $t_{2g}^4$ configurations in iridium ion. Filled and open circles stand for occupied and non-occupied electrons, respectively. $J$ ($j$) is multi-particle (single-particle) total angular momentum, respectively. The circles from left to right on each line represent $|m=-j,...,j\rangle$ states, respectively.    }
\end{figure}

In this paper, based on the multi-band $t$-$J$-like Hamiltonian derived from the $t_{2g}$ three-orbital Hubbard model in \nio, we investigate the dynamics of a single hole doped into the zigzag magnetic ordering state of \nio with internal degrees of freedom (IDF) , using the self-consistent Born approximation (SCBA). The IDF of the doped hole means that it does not only form a singlet but also triplet and quintet states.
According to the conservation of the total angular momentum $J$, we classify the $t_{2g}^4$ states, which are referred to as the IDF of the hole, into two singlet states, one triplet state, and two quintet states as illustrated in Fig.~\ref{fig1}(c). We find that the lowest-energy spectrum is dominated by the motion of the singlet $J=0$ hole and exhibits an electronlike dispersion (opening upward) near the $\Gamma$ point. The mid-energy and high-energy spectra are attributed to the motion of the triplet $J=1$ and quintet $J=2$ holes, respectively. The weight of the mid-energy spectra is concentrated on the $M$ point, consistent with the experiment observation\cite{Moreschini2017}. The high-energy spectra exhibit a holelike dispersion (opening downward) along the $\Gamma$-$K$ path. The overall features of spectra are robust, even when the Hund's coupling $J_H$ is changed from $0.17$ eV to $0.51$ eV, and are qualitatively irrelevant to the exchange model. The interference between IDF in different sublattices plays a significant role in inducing the complex dispersions.

In order to discuss the motion of a single hole in the  magnetically ordered background, $t$-$J$-like model is always used as the starting point\cite{PhysRevB.45.2425,PhysRevB.39.6880} and could be derived from multi-orbital Hubbard model by perturbation theory in the strong-coupling limit\cite{Yin2009} (also see Appendix \ref{ApdxA}):
\begin{align}\label{model}
  H_{tJ}=&P_0\sum_{ipp'}(\epsilon_{ip}\delta_{pp'}+\Delta_{i}^{pp'})X^{pp'}_iP_0\nonumber\\
  &+P_0\sum_{\mathclap{i<j\atop rr'ss'}}(V_{ij}^{rr',ss'}X_i^{rr'}X_j^{ss'}+h.c.)P_0+H_{\rm ex},
\end{align}
where $H_{\rm ex}$ contains the exchange interactions between the $J=1/2$ pseudospins. $P_0$ projects Hilbert space into the low-energy states including the $J=1/2$ states of $t^5_{2g}$ configuration and the $J=0,1,2$ states of $t^4_{2g}$ manifold as shown in Fig.~\ref{fig1}(c). $X_i^{pp'}=|ip\rangle\langle ip'|$, where $|ip\rangle$ satisfies $(H_{i}^U+H_i^{\lambda})|ip\rangle=\epsilon_{ip}|ip\rangle$, is a Hubbard operator on the $i$-th site. $H_i^{\lambda}=\lambda {\bm{l}}_i\cdot {\bm{s}}_i$ is the electron spin-orbit interaction. Here, $H_i^U$ contains intra-orbital Coulomb interaction $U$, inter-orbital Coulomb interaction $U^{\prime}$ , Hund's coupling $J_H$, and pairing hoping interaction  $J_P$ with  $J_P=J_H$ and $U=U^{\prime}+2J_H$. $\Delta_i^{pp'}=\langle ip|H_i^\Delta |ip'\rangle$ is the matrix element of the crystal field. $V_{ij}^{rr',ss'}=\sum_{\nu\nu'}t_{ij}^{\nu \nu'}\langle ir,js|c^\dagger_{i\nu}c_{j\nu'}|ir',js'\rangle$, where $c_{j,\nu'\equiv(l,\sigma)}$ annihilates an electron with spin $\sigma$ in the $l$-th orbital state at site $j$, are the parameters of hopping between the $J=1/2$ states and $t_{2g}^4$ states. $t_{ij}^{\nu\nu'}$ are the original hopping parameters between $t_{2g}$ orbitals.

To investigate the motion of a hole, we introduce hole fermions $f^\dagger_{is}$ and Schwinger bosons $b^\dagger_{is'}$ which correspond to the $t_{2g}^4$ and $J=1/2$ states (see Fig.~\ref{fig1}), respectively. Correspondingly, the electron operator in the low-energy space could be expressed by $c_{i\nu}=\sum_{ss'}\tilde{U}^{i\nu}_{ss'}f^\dagger_{is}b_{is'}$ with constraint $\sum_{s'}b^\dagger_{is'}b_{is'}+\sum_{s}f^\dag_{is}f_{is}=1$. Since the ground state of \nio is long-range zigzag ordering, the low-energy spin excitations are well described by the magnons. Within the linear spin-wave approximation (LSW), one Schwinger boson $b_{i0}$ is condensed and other bosons are mapped to Holstein-Primakoff bosons\cite{Muniz2014,Dong2018}. Using the Fourier transformation\cite{PhysRevB.73.155118,Sushkov1997}
 $f^\dag_{\bm{k}\alpha s} = \sqrt{4/N}\sum_{m}f^\dag_{m\alpha s}\mathrm{e}^{\mathrm{i}\bm{k}\cdot{\bm{r} _{m\alpha}}}$,
 where $N$ is the number of sites and $(m,\alpha)$ label the $\alpha$-th sublattice in the $m$-th magnetic unit cell,
 we represent the $t$-$J$-like model (Eq.~\ref{model}) in the matrix form within LSW,
 \begin{align}\label{hfb}
 H_{fb}=&\sum_{\mathclap{\bm{k}}}\hat{f}^\dag_{\bm{k}}\hat{h}(\bm{k})\hat{f}_{\bm{k}}+\sum_{\mathclap{\bm{q},n}}\omega_{n\bm{q}}\gamma_{n,\bm{q}}^\dagger\gamma_{n,\bm{q}}+\nonumber\\
 &\sum_{\bm{kq}n}\hat{f}^\dag_{\bm{k}}[\hat{g}_n(\bm{k},\bm{q})\gamma_{n,\bm{q}}
 +\hat{g}_n^\dag(\bm{k}\textrm{-}\bm{q},\textrm{-}\bm{q})\gamma^\dag_{n,\textrm{-}\bm{q}}]\hat{f}_{\bm{k}\textrm{-}\bm{q}},
 \end{align}
 where $\omega_{n\bm{q}}$ is the dispersion of the magnons $\gamma_{n,\bm{q}}^\dag$, $\hat{h}(\bm{k})$ is the Hamiltonian of the free hole, and $\hat{g}_n(\bm{k},\bm{q})$ describes the coupling between the $t_{2g}^4$ hole and the magnons. Here, $\hat{h}(\bm{k})$ and $\hat{g}_n(\bm{k},\bm{q})$ have a  matrix form, and the detailed derivations are given in Appendix \ref{ApdxA}.

\begin{figure}[tb]
  \centering
  \includegraphics[width=0.45\textwidth, trim={0.1cm 0.1cm 0.1cm 0.1cm}, clip]{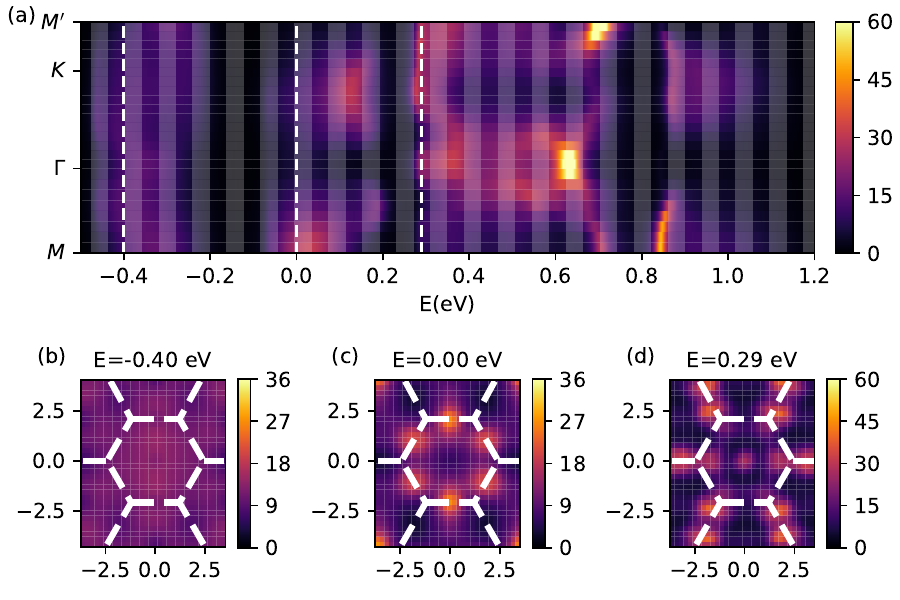}
  \caption{\label{fig2}(Color online) (a) Physical spectral function $A^c(\bm{k},\omega)$ along the high-symmetry path $M$-$\Gamma$-$K$-$M'$ as indicated in Fig.\ref{fig1}(b). The dashed lines denote the constant-energy cuts at $-0.40$ eV, $0.00$ eV, and $0.29$ eV as plotted in (b), (c), and (d) respectively. (b)-(d) Constant-energy contour maps of $A^c(\bm{k},\omega)$. The thick dashed lines denote Brillouin zone boundaries. }
\end{figure}

As the spectral properties measured in ARPES experiments are related to single-particle Green's function, we introduce two relevant Green's functions $\hat{G}^d(\bm{k},\omega)=\langle 0| \hat{d}_{\bm{k}}(\omega+\mathrm{i}\eta-H_{fb})^{-1}\hat{d}^\dag_{\bm{k}}|0\rangle$, where $|0\rangle$ is the ground state of the exchange term $H_{\rm ex}$ under the LSW and $d = f (c^{\dag})$ corresponds to the hole (physical) Green's function $\hat{G}^{f/c}(\bm{k},\omega)$, respectively. The hole Green's function is determined by the self-consistent equation under the SCBA\cite{PhysRevB.39.6880}. The electron operator $c_{\bm{k}\nu}=\sqrt{4/N}\sum_{i}c_{i\nu}\mathrm{e}^{\mathrm{i}\bm{k}\cdot{\bm{r}_i}}$ in the low-energy space could be expressed by hole operators and magnon operators
\begin{align}\label{copt1}
c_{\bm{k}\nu}=&\hat{f}^\dag_{\bm{k}}\hat{W}^{\nu}(\bm{k})+\sum_{\mathclap{\bm{q}n}}\hat{f}^\dag_{\bm{k}\textrm{-}\bm{q}}[\hat{Y}^{\nu}_n(\bm{k},\bm{q})\gamma_{n,\bm{q}}^\dag
+\hat{Y}^{\prime \nu}_n(\bm{k},\textrm{-}\bm{q})\gamma_{n,\text{-}\bm{q}}],
\end{align}
where $\hat{W}^\nu$ (see Appendix~\ref{ApdxA}) describes the creation amplitude of a single hole by moving an electron from zigzag state and $\hat{Y}^\nu_n$ ($\hat{Y}^{\prime \nu}_n$) (see Appendix~\ref{ApdxA}) denotes the creation (annihilation) amplitude of a magnon plus a single hole. With this, the physical Green's function can be calculated with the formula Eq.~\ref{gcgf}. The corresponding spectral functions for two Green's functions are defined as $A^f_{s}(\bm{k},\omega)=-\frac{1}{\pi}{\rm{Im}}\sum_{\alpha\beta}G^f_{\alpha s,\beta s}(\bm{k},\omega)$ and $A^c(\bm{k},\omega)=-\frac{1}{\pi}{\rm{Tr}}[{\rm{Im}}\hat{G}^c(\bm{k},\omega)]$.
In the following calculations, we choose $16\times 8$ magnetic unit cells (Fig.~\ref{fig1}(a)) and employ $U=1.7$ eV, $J_H=0.3$ eV, and $\lambda=0.4$ eV\cite{PhysRevB.93.214431}, unless otherwise specified. For the classical ground state of the exchange term $H_{\rm ex}$, we assume it is a zigzag order. The hopping parameters $t_{ij}^{\nu\nu'}$ and the crystal field $H^\Delta_i$ are from Ref.~\onlinecite{PhysRevB.93.214431}. The IDF are only considered up to the nine low-energy states, i.e., $J=0,1,2$ states (Fig.~\ref{fig1}(c)).

Let us first discuss the physical spectral function $A^c(\bm{k},\omega)$ which directly relates to the ARPES results, as shown in Fig.~\ref{fig2}. Figure~\ref{fig2}(a) shows that the low-energy spectrum (around $-0.4$ eV) has an electronlike dispersion near the $\Gamma$ point, which is well consistent with ARPES experiments\cite{Alidoust2016,Moreschini2017}. The corresponding contour map in Fig.~\ref{fig2}(b) exhibits maximum intensity near the $\Gamma$ point. In the mid-energy region between $-0.1$ eV and $0.2$ eV, the spectrum along the $\Gamma$-$K$ path hosts a downward parabolic dispersion and the intensities near the $\Gamma$ point are suppressed heavily. From the energy cut in Fig.~\ref{fig2}(c), we can see that the intensities of spectrum are concentrated around the $M$ point, which is consistent with the features of the 0.7-eV flat band in ARPES experiment\cite{Moreschini2017}. In the high-energy region between $0.4$ eV and $0.6$ eV, the intensities between $\Gamma$ point and $K$ point are suppressed and a holelike dispersion arises along the $\Gamma$-$K$ direction. Furthermore, the maximum intensities of spectrum at $0.29$ eV are shifted from near the $M$ point to near the $K$ point, relative to the the maximum intensities at $0.0$ eV, as shown in Fig.~\ref{fig2}(d). These features are also agreement with experiment\cite{Moreschini2017}.
{
In addition, our investigations have shown a similarity between the low-energy spectrum observed at the $\Gamma$ point and the entirety of the $\Gamma$-point spectrum documented in earlier studies\cite{Trousselet2013,Trousselet2014,Wang_2018}, where the incoherent high-energy spectral weight at the $\Gamma$ point aligns with the spectral weight near -0.3 eV in our study. However, it is worth highlighting that the intermediate and high-energy spectra exhibit distinct characteristics that deviate from those observed in prior investigations based on one-band $t$-$J$-like model\cite{Trousselet2013,Trousselet2014,Wang_2018}.
}

\begin{figure}[tb]
  \centering
  \includegraphics[width=0.45\textwidth, trim={0.5cm 0.2cm 2cm 1cm}, clip]{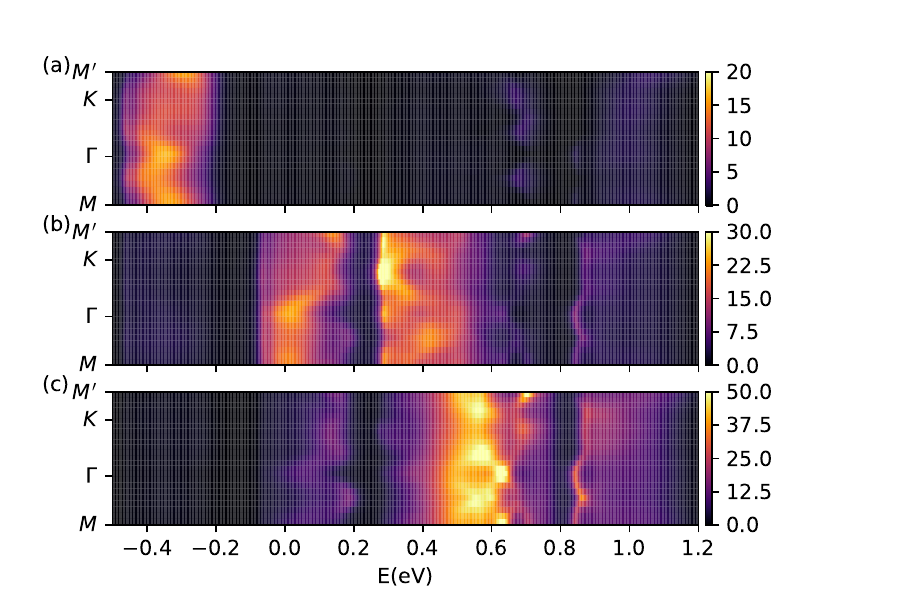}
  \caption{\label{fig3}(Color online) Spectral function $A^f_s(\bm{k},\omega)$ of a hole with IDF along the high-symmetry path $M$-$\Gamma$-$K$-$M'$ as indicated in Fig.\ref{fig1}(b). (a) Spectral function $A^f_0(\bm{k},\omega)$ of the $J=0$ hole. (b) Spectral function $A^{f,1}(\bm{k},\omega)$ of the $J=1$ hole. (c) Spectral function $A^{f,2}(\bm{k},\omega)$ of the $J=2$ hole. }
\end{figure}

To understand the contribution of the hole with different IDF to the physical spectral function, we turn to the hole spectral function $A^f_s(\bm{k},\omega)$ with different total angular momenta. Figure~\ref{fig3}(a) shows the spectral function $A^f_0(\bm{k},\omega)$ of the hole with total angular momentum $J=0$. The spectral intensities of $A^f_0(\bm{k},\omega)$ are mainly concentrated in the low-energy region. This suggests that the low-energy spectra in $A^c(\bm{k},\omega)$ come basically from the motion of the singlet $J=0$ hole.
Figure~\ref{fig3}(b) shows the spectral function $A^{f,1}({\bm{k},\omega})=\sum_{s=-1}^{1}A^f_s(\bm{k},\omega)$ of the triplet $J=1$ hole. The weight of the triplet-hole spectra are concentrated in the mid-energy region, indicating that the mid-energy spectra in $A^c(\bm{k},\omega)$ (see Fig.~\ref{fig2}) are dominated by the motion of the triplet $J=1$ hole, although there is also a contribution from the motion of the $J=2$ hole as shown in Fig.~\ref{fig3}(c), where the spectral function $A^{f,2}({\bm{k},\omega})=\sum_{s=-2}^{2}A^f_s(\bm{k},\omega)$ of the quintet $J=2$ hole is exhibited.  From the mid-energy spectra in Fig.~\ref{fig3}(b) and (c), we find that the intensities near the $\Gamma$ point are not suppressed, on contrary to the spectra in the physical spectral function $A^c(\bm{k},\omega)$ of Fig.~\ref{fig2}.
This is due to the absence of the vertex $\hat{W}$ that would give rise to the interference between IDF in different sublattices.
\begin{figure}[tb]
  \centering
  \includegraphics[width=0.45\textwidth, trim={ 1cm 0.2cm 2cm 0.5cm}, clip]{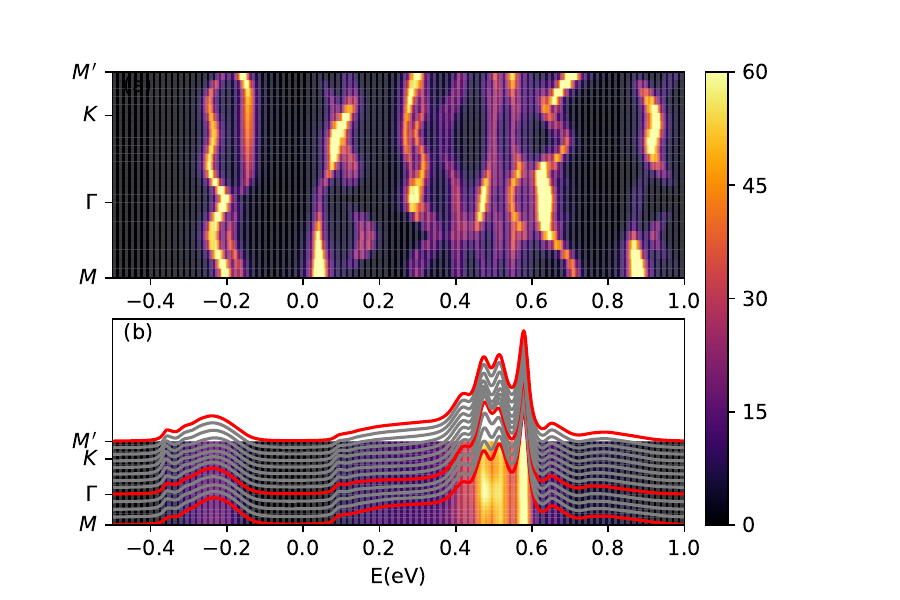}
  \caption{\label{fig4}Physical spectral functions $A^c(\bm{k},\omega)$ for (a) the nearly free hole  and (b) the only-scattered hole (see text).}
\end{figure}

To gain more insights into the scattering process, we present the physical spectral functions induced by the nearly free hole (by setting the coupling $\hat{g}_n(\bm{k},\bm{q})=0$) and only-scattered hole (by setting  $\hat{h}(\bm{k})=0$ except the on-site terms) in Fig.~\ref{fig4}(a) and (b), respectively. Figure~\ref{fig4}(a) shows that the dispersion of the lowest-energy band near the $\Gamma$ point is electronlike, and a near flat band with strongly M-point intensity appears at mid-energy region around 0.02 eV. In Fig.~\ref{fig4}(b), it is clearly shown that the spectrum takes the shape of a ladder. This feature implies that the hole moving in the zigzag order background is strongly scattered by the magnons, which is similar to the string excitation in a N\'{e}el spin background\cite{PhysRevB.45.2425}. Compared with Fig.~\ref{fig4}(a), the bands in Fig.~\ref{fig2}(a) are greatly broadened and heavily renormalized, but the main features, including electronlike dispersion near the $\Gamma$ point and high M-point intensity of mid-energy states, remain.

\begin{figure}[tb]
  \centering
  \includegraphics[width=0.45\textwidth, trim={ 0.5cm 0.2cm 2cm 0.5cm}, clip]{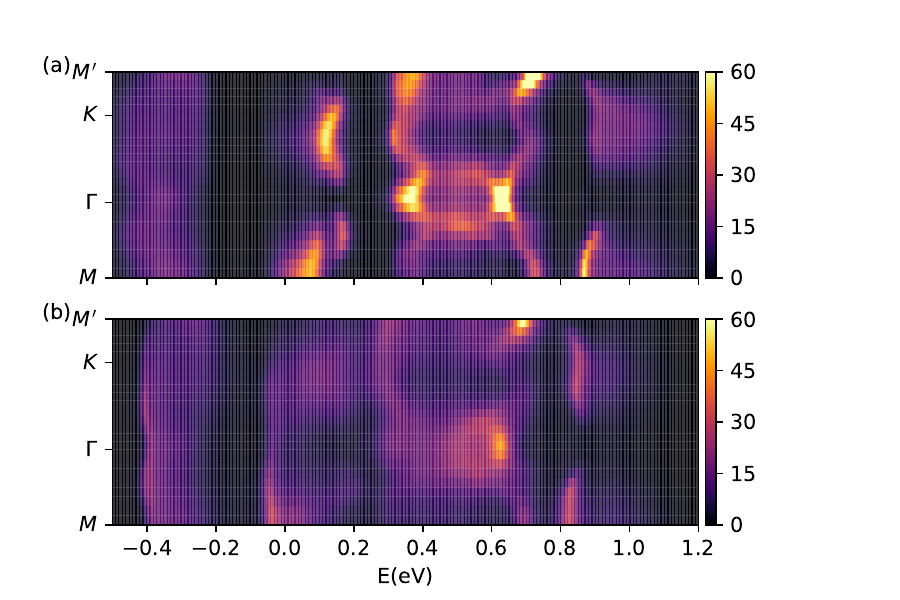}
  \caption{\label{fig5} (a) and (b) Physical spectral function $A^c(\bm{k},\omega)$ for Hund's coupling $J_H=0.17$ eV and $0.51$ eV, respectively. }
\end{figure}

It is noteworthy that the main features mentioned above remain robust even when the Hund's coupling $J_H$ is altered, as depicted in Fig.~\ref{fig5}(a) and (b), where $J_H$ is changed from $0.3$ eV to $0.17$ eV and $0.51$ eV, respectively. As $J_H$ increases, the gap between the singlet and triplet hole (near $-0.2$ eV) decreases, and both the low-energy dispersion near the $\Gamma$ point and the mid-energy dispersion near the M point become flatter (Fig.~\ref{fig5}(b)).
The reduction of the gap is mainly due to the decreasing difference of the on-site energy between the singlet and triplet states of the hole, as $J_H$ increase, i.e., $\epsilon_{J=1}{-}\epsilon_{J=0}=(\sqrt{25J^2_H+10\lambda J_H+9\lambda^2}-5J_H)/2$. At the mid-energy region, the increase of $J_H$ causes the spectra shown in Fig.~\ref{fig5}(b) to extend further, resulting in a reduction of the gap $\Delta_1$ (near $0.2$ eV). The holelike dispersion along the $\Gamma$-$K$ path at high-energy region still remains with the increase of $J_H$. Moreover, these overall features are also observed (see Appendix~\ref{ApdxB}) when we replace the exchange term $H_{\rm ex}$ in Eq.~\ref{model} with the exchange model obtained from the exact diagonalization method \cite{PhysRevB.93.214431}.

Based on the physical and hole spectral function, we suggest that the ARPES spectra at low, medium, and high energies are predominantly influenced by the singlet, triplet, and quintet states, respectively. Our findings qualitatively reproduce the key characteristics of the ARPES spectra, including (i) an electron-like dispersion near the $\Gamma$ point at low energies, (ii) a flat band with spectral weights concentrated on the $M$ point, and (iii) a high-energy hole-like dispersion along the $\Gamma$-$K$ path. These agreements with experimental observations suggest that \nio is a spin-orbit Mott insulator.
{
Quantitatively, the energy gap near $-0.2$ eV and the gap $\Delta_1$ are not clearly observed in ARPES experiments\cite{PhysRevLett.109.266406,Alidoust2016,Moreschini2017,PhysRevB.101.235415}. The low-energy spectrum, instead of being concentrated near the $\Gamma$ point as expected, disperses throughout the Brillouin zone. This is inconsistent with experimental findings\cite{Alidoust2016,Moreschini2017,PhysRevB.101.235415}. This inconsistency may arise from neglecting the matrix-element effect and the mixing between $j=1/2$ and $3/2$ states. Additionally, the high-energy spectra above 0.7 eV also deviate from ARPES results due to the neglect of the high-energy IDF, as illustrated in Fig.~\ref{fig1}(c).
}
In contrast to previous theoretical works on \nio\cite{Trousselet2013,Trousselet2014,Wang_2018}, our study takes into account more IDF of the hole. In the previous works, the hole corresponds to the missing $J=1/2$ state, which is associated only with the singlet state in our current research.
In contrast to the similar study on Sr$_2$IrO$_4$\cite{Paerschke2017}, we find that in \nio, the spectra from the quintet and triplet states overlap heavily, and the interference between IDF in different sublattices plays a significant role in producing the dispersion at medium and high energies.

In summary, we study the dynamics of a hole with IDF under the zigzag magnetic background in \nio and explain various features in the ARPES spectra using the SCBA. We find that the electronlike dispersive states in the low-energy region arise from the singlet hole, while the flat band with maximum $M$-point intensities in the mid-energy region is attributed to the motion of the triplet hole. The holelike dispersive spectra in the high-energy region are due to the motion of the quintet hole. The interference between the IDF of hole in different sublattices plays a significant role in producing the complex dispersions. However, to accurately reproduce the fine structure of the ARPES spectra, additional information such as the matrix-element effects and the mixing between the $j=1/2$ and $3/2$ states may need to be considered.


This work was supported by National Key Projects for Research and Development of China (Grant No. 2021YFA1400400), National Natural Science Foundation of China (Grant No. 12004191, 92165205 and 12074175), Natural Science Foundation of Jiangsu Province (Grant No. BK20200738), and the Natural Science Research Start-up Foundation of Posts and Telecommunications (Grant No. NY220095).

\appendix
\section{\uppercase{Detail derivation}}\label{ApdxA}
The $t$-$J$-like model originates from the effective description of the low-energy space of the Hubbard model in the strong-coupling limit. The multi-orbital Hubbard model is given by
\begin{align}\label{hubbardmodel}
H=&\sum_{\mathclap{i\neq j\nu\nu'}}t_{ij}^{\nu\nu'}c^\dag_{i\nu}c_{j\nu'}+\sum_{i\nu\nu'}\Delta_{i\nu\nu'}c^\dag_{i\nu}c_{i\nu'} +\lambda \sum_i\bm{l}_i\cdot \bm{s}_i
+ H_U,
\end{align}
where the first three terms are hopping term $H^t$, crystal field term $H^\Delta$, and spin-orbit term $H^{\lambda}$, respectively. Here, $H_t$ contains up to the third nearest-neighbor hopping term. $H_U$ is Hubbard interaction term which is written as
\begin{align}
H_U&=\dfrac{U}{2}\sum_{\mathclap{il\sigma\neq\sigma'}}c^\dag_{il\sigma}c^\dag_{il\sigma'}c_{il\sigma'}c_{il\sigma}
+\dfrac{U'}{2}\sum_{\mathclap{il\neq l'\sigma\sigma'}}c^\dag_{il\sigma}c^\dag_{il'\sigma'}c_{il'\sigma'}c_{il\sigma}\nonumber \\
&+\dfrac{J_H}{2}\sum_{\mathclap{il\neq l'\sigma\sigma'}}c^\dag_{il\sigma}c^\dag_{il'\sigma'}c_{il\sigma'}c_{il'\sigma}
+\dfrac{J_P}{2}\sum_{\mathclap{il\neq l'\sigma\neq\sigma'}}c^\dag_{il\sigma}c^\dag_{il\sigma'}c_{il'\sigma'}c_{il'\sigma}.
\end{align}
To reduce the multi-orbital Hubbard model to the effective $t$-$J$-like model, we introduce projection operators $P_1=1-P_0$ and $P_0=\prod_i\sum_{p\in g.s.} X_i^{pp}$ where $g.s.$ means the low-energy eigenstate of $H_U+H^\lambda$. The Hubbard operators at site $i$ and $j$ satisfy the commutation relations
\begin{equation}\label{xoperator}
  \left[X_i^{pp'},X_j^{ss'}\right]_{\pm}=\delta_{ij}(X_i^{ps'}\delta_{p's}\pm X_i^{sp'}\delta_{ps'}),
\end{equation}
where the plus sign is used only if both Hubbard operators belong to Fermi type, i.e. change the number of electrons by an odd number. The minus sign corresponds to boson-type Hubbard operators. Any operator $\hat{O}_i$ on a site $i$ can be expressed in terms of Hubbard operators as
\begin{align}
\hat{O}_i=\sum_{pp'}\langle ip|\hat{O}_i|ip'\rangle X_i^{pp'}.
\end{align}

Following the Ref.~\onlinecite{Yin2009}, we define Hamiltonian $H=H_0+H_1$, where $H_0=P_0 H P_0+P_1 H P_1$ is an unperturbed Hamiltonian and $H_1=P_1 H P_0+P_0 H P_1$ is a perturbation term. After approximating to second order, we obtain
\begin{align}
H_{\rm eff}=H_0+\frac{1}{2}[H_1,S],
\end{align}
where
\begin{align}
H_0=&\sum_i\sum_{{pp'\in g.s.}}(\epsilon_{ip}\delta_{pp'}+\Delta_i^{pp'})X_i^{pp'}\nonumber \\
&+P_0\sum_{\mathclap{i<j\atop rr'ss'}}\left(V_{ij}^{rr',ss'}X_i^{rr'}X_j^{ss'}+H.c.\right)P_0, \\
H_1=&P_0\sum_{\mathclap{i<j\atop rr'ss'}}\left(V_{ij}^{rr',ss'}X_i^{rr'}X_j^{ss'}+H.c.\right)P_1+H.c.,\\
S=&P_0\sum_{\mathclap{i<j\atop rr'ss'}}\left(\dfrac{V_{ij}^{rr',ss'}X_i^{rr'}X_j^{ss'}}{\epsilon_{r'}+\epsilon_{s'}-\epsilon_{r}-\epsilon_{s}}-H.c.\right)P_1 -H.c..
\end{align}
Here we neglect the effect of the crystal field $H^\Delta$ on the second-order Hamiltonian, i.e., $\frac{1}{2}[H_1,S]$. The second-order Hamiltonian contains two-site exchange term and three-site hopping term. Since the hopping amplitude in the second-order Hamiltonian is great smaller than that in the zero-order Hamiltonian $H_0$, the three-site hopping term is ignored in our calculation. The two-site exchange terms can be formally written as
\begin{align}
H_{\rm{ex}}=&\sum_{i<j}(J_{ij,M}^{rr'ss'}X_{i,M}^{rr'}X_{j,M}^{ss'}+J_{ij,M\textrm{-}1}^{rr'ss'}X_{i,M\textrm{-}1}^{rr'}X_{j,M\textrm{-}1}^{ss'}+\nonumber\\
&J_{ij,M,M\textrm{-}1}^{rr'ss'}X_{i,M}^{rr'}X_{j,M\textrm{-}1}^{ss'}+J_{ij,M\textrm{-}1,M}^{ss'rr'}X_{i,M\textrm{-}1}^{ss'}X_{j,M}^{rr'}),
\end{align}
where the repeated superscript $r$,$r'$,$s$,$s'$ is summed up, $X_{i,M}^{rr'}$ is a boson-type Hubbard operator, and subscript $M$ indicates the number of electrons in the state $r$ or $r'$. In one-hole case, we only keep the first term of $H_{\rm ex}$. Here, the exchange interactions are also maintained up to the third nearest-neighbor term.

\begin{figure}[tb]
  \centering
  \includegraphics[width=0.45\textwidth, trim={ 0.1cm 0.2cm 0.1cm 0.1cm}, clip]{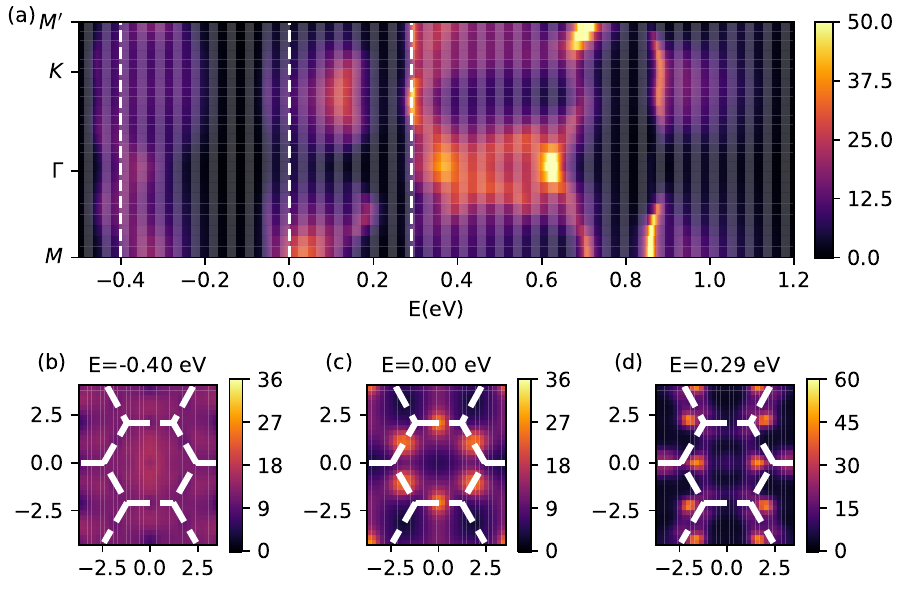}
  \caption{\label{fig6}Physical spectral function $A^c(\bm{k},\omega)$ for the exchange model obtained by the exact diagonalization method. (a) Physical spectral function $A^c(\bm{k},\omega)$ along the high-symmetry path $M$-$\Gamma$-$K$-$M'$. The dashed lines denote the constant-energy cuts at $-0.40$ eV, $0.00$ eV, and $0.29$ eV as plotted in (b), (c), and (d) respectively. (b)-(d) Constant-energy contour maps of $A^c(\bm{k},\omega)$. Thick dashed lines denote Brillouin zone boundaries.  }
\end{figure}

Due to the background of the magnetic order, we map the $J=1/2$ states and $t_{2g}^4$ states into Schwinger bosons $\tilde{b}_{ir}$ and hole fermions $f_{is}$, respectively.
Thus the effective Hamiltonian reads
\begin{align}\label{hbf0}
H_{\rm eff}=&\sum_{is}(\epsilon_{is}\delta_{ss'}+\Delta_{i}^{ss'} )f^\dag_{is}f_{is'}+\sum_{ir}(\epsilon_{ir}\delta_{rr'}+\Delta_{i}^{rr'} )\tilde{b}^\dag_{ir}\tilde{b}_{ir'}\nonumber \\
&+\sum_{i<j\atop rr'ss'}\left(V_{ij}^{rr',ss'}\tilde{b}^\dag_{ir}f_{ir'}f^\dag_{js}\tilde{b}_{js'}+H.c.\right) \nonumber \\
&+ \sum_{i<j\atop rr'ss'}J_{ij,N}^{rr'ss'}\tilde{b}^\dag_{ir}\tilde{b}_{ir'}\tilde{b}^\dag_{js}\tilde{b}_{js'},
\end{align}
where the second term will be omitted because $\hat{\epsilon}_{i}$ and $\hat{\Delta}_{i}$ are constant matrices for the $J=1/2$ states.
To obtain the ordered spin state, we take a unitary transformation $\tilde{b}^\dag_{ir}=\sum_{s'}U^{\prime*}_{i,rs'}b^\dag_{is'}$, where $b^\dag_{i0}$ creates the ordered spin state and is replaced by a number\cite{Dong2018} according to the constraint $\sum_{s'}b^\dagger_{is'}b_{is'}+\sum_{s}f^\dag_{is}f_{is}=1$. Substituting the unitary transformation into the Eq.~\ref{hbf0}, we get
\begin{align}
H_{fb}\simeq&\sum_{is}(\epsilon_{is}\delta_{ss'}+\Delta_{i}^{ss'} )f^\dag_{is}f_{is'}+\sum_{\mathclap{i<j\atop rr'ss'\neq0}}(\tilde{V}_{ij}^{0r',s0}f_{ir'}f^\dag_{js}\nonumber\\
&+\tilde{V}_{ij}^{0r',ss'}f_{ir'}f^\dag_{js}b_{js'}+\tilde{V}_{ij}^{rr',s0}b^\dag_{ir}f_{ir'}f^\dag_{js}+H.c.)\nonumber\\
&+H_{\rm LSW},
\end{align}
where $b_{ir\neq0}^\dag$ is Holstein-Primakoff boson and $H_{\rm LSW}$ is the magnon Hamiltonian under LSW. After Fourier transformation, the Hamiltonian $H_{fb}$ is written as Eq.~\ref{hfb}, where $\hat{h}(\bm{k})$ and $\hat{g}_n(\bm{k},\bm{q})$ are given by
\begin{align}
h(\bm{k})_{\alpha s,\beta r'} =&[ \epsilon_{\alpha s}\delta_{\alpha\beta}\delta_{sr'}+\Delta^{sr'}_{\alpha}\delta_{\alpha\beta}-\nonumber\\
&\sum_{j\textrm{-}i}\tilde{V}_{\beta\alpha}^{0r',s0}(j\textrm{-}i)\mathrm{e}^{\textrm{-}\mathrm{i}\bm{k}\cdot(\bm{r}_j\textrm{-}\bm{r}_i)}]
\end{align}
and
\begin{align}
g_n^{{\beta s,\alpha r'}}(\bm{k},\bm{q})=&\dfrac{-1}{\sqrt{N/4}}\sum_{j\textrm{-}i}\sum_{s'\neq0}[\tilde{V}_{\alpha\beta}^{0r',ss'}(j\textrm{-}i)u_{\beta s',n}(\bm{q})\nonumber\\
&\times \mathrm{e}^{\textrm{-}\mathrm{i}(\bm{k}\textrm{-}\bm{q})\cdot(\bm{r}_j\textrm{-}\bm{r}_i)}
+\tilde{V}_{\alpha\beta}^{s'r',s0}(j\textrm{-}i)v_{\alpha s',n}(\bm{q})\nonumber\\
&\times \mathrm{e}^{\textrm{-}\mathrm{i}\bm{k}\cdot(\bm{r}_j\textrm{-}\bm{r}_i)}],
\end{align}
respectively. $u_{\beta s',n}(\bm{q})$ and $v_{\alpha s',n}(\bm{q})$ are the components of the eigenvector of $H_{\rm LSW}$ and meet
\begin{align}\label{bgamma}
b_{\bm{q}\beta s}=\sum_n u_{\beta s,n}(\bm q)\gamma_{n,\bm{q}}+v^{\ast}_{\beta s,n}(\textrm{-}\bm{q})\gamma_{n,\textrm{-}\bm{q}}^\dag.
\end{align}

In the SCBA, the self-consistent equation of the hole Green's function is expressed by
\begin{align}
\hat{G}^f(\bm{k},\omega)=&[\omega+\mathrm{i}\eta-\hat{h}(\bm{k})-\hat{\Sigma}(\bm{k},\omega)]^{-1},
\end{align}
where
\begin{align}
\hat{\Sigma}(\bm{k},\omega)=\sum_{n\bm{q}}\hat{g}_n(\bm{k},\bm{q})\hat{G}^f(\bm{k}\textrm{-}\bm{q},\omega\textrm{-}\omega_{n\bm{q}})\hat{g}^\dag_n(\bm{k},\bm{q}).
\end{align}
Here, we choose $\eta=0.01$ and set 2000 points from -2.0 eV to 2.5 eV in the $\omega$ mesh.

Since an electron removed from different sublattices in ARPES experiment can not be distinguished, it is more reasonable to compare the imaginary part of the physical Green's function $\hat{G}^c$ with the ARPES spectrum\cite{Sushkov1997}.
Then the electron operator $c_{i\nu}$ acting in the low-energy space is given by
\begin{align}
c_{i\nu} = \sum_{sr}\langle s|c_{i\nu}|r\rangle X_i^{sr}
         = \sum_{sr} {U}_{sr}^{i\nu}f_{is}^\dag \tilde{b}_{ir}=\sum_{ss'} \tilde{U}_{ss'}^{i\nu}f_{is}^\dag {b}_{is'},
\end{align}
where $\tilde{U}_{ss'}^{i\nu}=\sum_r U_{sr}^{i\nu}U_{i,rs'}^{\prime}$. In the momentum space, the electron operator is
\begin{align}\label{copt2}
c_{\bm{k}\nu}=\sum_{\mathclap{\beta s}}(f^\dag_{\bm{k}\beta s}\tilde{U}_{s0}^{\beta\nu}\bar{b}_{\beta 0}+\sqrt{\frac{4}{N}}\sum_{\mathclap{\bm{q}s'\neq0}}f^\dag_{\bm{k}\textrm{-}\bm{q}\beta s}\tilde{U}^{\beta\nu}_{ss'}b_{\textrm{-}\bm{q}\beta s'})\mathrm{e}^{\mathrm{i}\bm{k}\cdot\bm{r}_{\beta}},
\end{align}
where $\bar{b}_{\beta0}$ denotes the amplitude of the magnetic moment and is equal to $\sqrt{1\textrm{-}4/N\sum_{\bm{q}s\neq0}\langle b^\dag_{\bm{q}\beta s}b_{\bm{q}\beta s}\rangle}$. Substituting Eq.~\ref{bgamma} into Eq.~\ref{copt2}, we obtain the Eq.~\ref{copt1} in the main text, where
\begin{align}
W_{\beta s}^{\nu}(\bm{k})&=\tilde{U}_{s0}^{\beta\nu}\sqrt{1-\frac{4}{N}\sum_{s'\bm{q}n}|v_{\beta s',n}(\bm{q})|^2}\mathrm{e}^{\mathrm{i}\bm{k}\cdot r_{\beta}},\nonumber\\
Y^\nu_{\beta s,n}(\bm{k},\bm{q})&=\sqrt{\frac{4}{N}}\sum_{s'\neq0}\tilde{U}_{ss'}^{\beta\nu}v_{\beta s',n}^*(\bm{q})\mathrm{e}^{\mathrm{i}\bm{k}\cdot \bm{r}_\beta},\nonumber
\end{align}
and
\begin{align}
Y^{\prime \nu}_{\beta s,n}(\bm{k},\textrm{-}\bm{q})=\sqrt{\frac{4}{N}}\sum_{s'\neq0}\tilde{U}_{ss'}^{\beta\nu}u_{\beta s',n}(\textrm{-}\bm{q})\mathrm{e}^{\mathrm{i}\bm{k}\cdot \bm{r}_\beta}.
\end{align}
Following the approximation of Ref.~\onlinecite{Sushkov1997}, the physical Green's function is given by
\begin{align}\label{gcgf}
&\hat{G}^c(\bm{k},\omega)=\hat{W}^\dag(\bm{k})\hat{G}^{f}(\bm{k},\omega)\hat{W}(\bm{k})\nonumber\\
&+\sum_{\mathclap{n\bm{q}}}\hat{Y}^\dag_n(\bm{k},\bm{q})\hat{G}^f(\bm{k}\textrm{-}\bm{q},\omega\textrm{-}\omega_{n\bm{q}})\hat{Y}_n(\bm{k},\bm{q})\nonumber\\
&+\sum_{n\bm{q}}\hat{Y}^\dag_n(\bm{k},\bm{q})\hat{G}^f(\bm{k}\textrm{-}\bm{q},\omega\textrm{-}\omega_{n\bm{q}})\hat{g}^\dag_n(\bm{k},\bm{q})\hat{G}^f(\bm{k},\omega)\hat{W}(\bm{k})\nonumber\\
&+\sum_{n\bm{q}}\hat{W}^\dag(\bm{k})\hat{G}^f(\bm{k},\omega)\hat{g}_n(\bm{k},\bm{q})\hat{G}^f(\bm{k}\textrm{-}\bm{q},\omega\textrm{-}\omega_{n\bm{q}})\hat{Y}_n(\bm{k},\bm{q})\nonumber\\
&+\sum_{\mathclap{nn'\bm{q}\bm{q}^\prime}}\hat{Y}^\dag_n(\bm{k},\bm{q})\hat{G}^f(\bm{k}\textrm{-}\bm{q},\omega\textrm{-}\omega_{n\bm{q}})\hat{g}_n^\dag(\bm{k},\bm{q})\hat{G}^f(\bm{k},\omega)\nonumber\\
&\hat{g}_{n'}(\bm{k},\bm{q}^\prime)\hat{G}^f(\bm{k}\textrm{-}\bm{q}^\prime,\omega\textrm{-}\omega_{n'\bm{q}^\prime})\hat{Y}_{n'}(\bm{k},\bm{q}^\prime).
\end{align}
Since the ground state $|0\rangle$ is approximate to the ground state of $H_{\rm LSW}$, the annihilation of a magnon in the ground state is zero, i.e. $\gamma_{n,\textrm{-}\bm{q}}|0\rangle=0$. Thus, the vertex $\hat{Y}^{\prime\nu}$ does not appear in Eq.~\ref{gcgf}.

\section{\uppercase{spectra with different exchange models}}\label{ApdxB}

In the main text, the exchange model obtained by the perturbation theory is varied with the change of Hund's coupling from 0.17 eV to 0.51 eV. In this process, we have assumed that the ground state of the exchange model is the zigzag order, and verified that the zigzag order is the local minimum of the zero-th exchange model within the spin-wave theory\cite{Wang2017,Dong2018}.

The exchange interactions in $H_{\rm{ex}}$ are also obtained by the six-site exact diagonalization method\cite{PhysRevB.93.214431}. The values of the exchange interactions are highlighted in bold in the Table II of Ref.~\onlinecite{PhysRevB.93.214431}. Even though the crystal field is not considered in the six-site exact diagonalization method, it is used in the $t$-$J$-like model (Eq.~\ref{hbf0}), where the on-site hopping terms $\Delta_i^{ss'}$ is induced by the crystal field. The physical spectral function $A^c(\bm{k},\omega)$ for this case is shown in Fig.~\ref{fig6}. The spectra are almost the same as that in Fig.~\ref{fig2} of the main text. It implies that the features of the spectra are robust and irrelevant to the exchange model.

\bibliography{paper}
\end{document}